\documentclass[sigconf,letterr]{acmart}

\usepackage[absolute,showboxes]{textpos}
\usepackage[utf8]{inputenc}
\usepackage{booktabs}
\usepackage{subcaption}
\usepackage[inline]{enumitem}
\usepackage{standalone}
\usepackage{tikz}
\usetikzlibrary{calc,colorbrewer,decorations,decorations.pathreplacing,fit}

\setcopyright{none}
\copyrightyear{2019}
\acmYear{2019}
\acmConference[ICN '19]{6th ACM Conference on Information-Centric Networking}{September 24--26, 2019}{Macau, Macao}
\acmBooktitle{6th ACM Conference on Information-Centric Networking (ICN '19), September 24--26, 2019, Macau, Macao}
\acmPrice{15.00}
\acmDOI{10.1145/3357150.3357398}
\acmISBN{978-1-4503-6970-1/19/09}

\renewcommand\footnotetextcopyrightpermission[1]{}

\mathchardef\UrlBreakPenalty=10000

\newenvironment{italicquotes}
{\begin{quote}\itshape}
{\end{quote}}

\usepackage{balance}

\usepackage{pifont}

\usepackage{xspace}

\newcommand{\ie}{\textit{i.e.,}~}
\newcommand{\etc}{\textit{etc.}~}
\newcommand{\one}{({\em i})\xspace}
\newcommand{\two}{({\em ii})\xspace}
\newcommand{\three}{({\em iii})\xspace}

\makeatletter
\renewcommand{\paragraph}[1]{\vspace*{0.03in}\noindent{\bf #1.}\hspace{0.25ex \@plus1ex \@minus.2ex}}
\makeatother

\graphicspath{{figs/}}

\begin{document}
\title[BT Mesh: How much ICN is Inside?]{Bluetooth Mesh under the Microscope:\\ How much ICN is Inside?}

\author{Hauke Petersen}
\affiliation{%
  \institution{Freie Universit\"at Berlin}
}
\email{hauke.petersen@fu-berlin.de}

\author{Peter Kietzmann}
\affiliation{%
  \institution{HAW Hamburg}
}
\email{peter.kietzmann@haw-hamburg.de}

\author{Cenk G{\"u}ndo\u{g}an}
\affiliation{%
  \institution{HAW Hamburg}
}
\email{cenk.guendogan@haw-hamburg.de}

\author{Thomas C. Schmidt}
\affiliation{%
  \institution{HAW Hamburg}
}
\email{t.schmidt@haw-hamburg.de}

\author{Matthias W{\"a}hlisch}
\affiliation{%
  \institution{Freie Universit\"at Berlin}
}
\email{m.waehlisch@fu-berlin.de}

\renewcommand{\shortauthors}{H. Petersen et al.}

\begin{abstract}
	Bluetooth (BT) mesh is a new mode of BT operation for low-energy devices that offers group-based publish-subscribe as a network service with additional caching capabilities. These features resemble concepts of information-centric networking (ICN), and the analogy to ICN has been repeatedly drawn in the BT community. In this paper, we compare BT mesh with ICN both conceptually and in real-world experiments. We contrast both architectures and their design decisions in detail. Experiments are performed on an IoT testbed using NDN/CCNx and BT mesh on constrained RIOT nodes. Our findings indicate significant differences both in concepts and in real-world performance. Supported by new insights, we identify synergies and sketch a design of a BT-ICN that benefits from both worlds.
\end{abstract}


\maketitle

\setlength{\TPHorizModule}{\paperwidth}
\setlength{\TPVertModule}{\paperheight}
\TPMargin{5pt}
\begin{textblock}{0.8}(0.1,0.02)
     \noindent
     \footnotesize
     If you cite this paper, please use the ICN reference:
     H. Petersen, P. Kietzmann, C. G{\"u}ndo\u{g}an, T.~C. Schmidt, M. W\"ahlisch. Bluetooth Mesh under the Microscope: How much ICN is Inside?. In \emph{Proc. of ACM ICN}, ACM, 2019.
\end{textblock}

\section{Introduction}\label{sec:intro}

\begin{italicquotes}
The Bluetooth mesh standard adopted last year is a wireless technology
based on the ICN principles. [\dots] It implements all of the major
paradigms of information-centric networking in order to enable simplicity,
scalability and reliability [\dots].
\rule[0.5ex]{1.5em}{0.5pt}~\cite{s-icnra-18}
\end{italicquotes}

The above claim triggered the investigations of the present paper. We are curious how information-centric networking (ICN) features re-appear in  Bluetooth (BT) mesh, a new flavor of Bluetooth Low Energy (BLE) that was designed for intelligent home installations such as smart lighting, and how they compare to the  ICN  instances NDN/CCNx \cite{jstp-nnc-09,zabjc-ndn-14}.

In recent years, BLE has seen a remarkable adoption for the sake of introducing  cheap and popular personal area networking (PAN) to the constrained embedded world. BLE promises easy and robust networking between sensors or actuators and mobile devices such as smartphones, tablets, notebooks.
BLE offers two modes of operation: a connection-oriented point-to-point mode and a connection-less advertising mode.
BT mesh is a new flavor of BLE communication, based on the latter.
It supports multi-hop networking, network layer caching, and a group-based publish-subscribe, which resembles ICN features---but is BT mesh really built according to ICN principles?

We investigate this question from two perspectives. First, we explore and discuss BT networking functions in comparison with ICN principles and their concrete implementation in NDN/CCNx. Second, we empirically contrast the technologies by performing common ICN and BLE experiments on RIOT \cite{bhgws-rotoi-13,bghkl-rosos-18} in a testbed deploying CCN-lite and the NimBLE stack.

 Our findings from basic experiments measuring  standard metrics indicate that in simple scenarios BT mesh performs comparably to NDN at the price of consuming significantly higher network resources. In more complex multi-hop settings, NDN easily outperforms BT mesh due to its more efficient packet transport and more widely applicable in-network caching. These insights obtained from conceptual review and experiments lead us to outline a future approach of how BT and ICN technologies could converge to a truly information-centric low-power wireless PAN (LoWPAN) system.

The remainder of this paper is structured as follows. We first dive into the BT mesh technology in Section \ref{sec:blemesh}, followed by a detailed theoretical comparison of ICN and NDN concepts in Section \ref{sec:icn-ble}. Experimental evaluations are discussed in Section \ref{sec:eval}. Section \ref{sec:perspect} presents our technological vision on future ICN networking based on Bluetooth. Related work on BT and NDN performance is supplied in Section \ref{sec:related-work}.

%%% Local Variables:
%%% mode: latex
%%% TeX-master: "main"
%%% End:

%\vspace{-1.175cm}
\section{Bluetooth Mesh}\label{sec:blemesh}
BLE~\cite{b-bcsv-19} focuses on robust wireless communication in the open 2.4~GHz band and is widely used in IoT scenarios due to its omnipresence on mobiles.
BT mesh~\cite{b-bmpsv-19} is standardized on top of BLE, offering management of multi-hop topologies and enabling many-to-one and many-to-many group communication.

The BT mesh standard specifies a complete networking system, starting from the mapping to BLE as transport over network configurations including functional roles of participants and the provisioning of security credentials, up to  application layer formats.
In the following, we focus on aspects of BT mesh that are most relevant when comparing with ICN.
For this, we do not further cover everything related to node provisioning or the proxy role.

\subsection{Basic communication primitives}

\paragraph{Lower layers}
A Bluetooth mesh node always connects to three broadcast channels for advertisement.
Data is transmitted consecutively to all of these channels in a defined advertisement interval.
Receivers constantly scan for packets on all three channels alternately, leading to a 100~\% duty-cycle. This sacrifices  energy savings attained by BLE.

\paragraph{Network layer}
The network layer defines four types of addresses: unassigned, unicast, group, and virtual. Group addresses are fixed and map to different roles of a node in the mesh network, \ie \textit{all-relays} and \textit{all nodes}. Virtual addresses act as configurable multicast group addresses, which are mainly set by application models during network provisioning time, but it is also possible to define virtual addresses at runtime using distributed control methods.

\paragraph{Application layer}
Addressing data for applications is organized by a \textit{model layer} that specifies different application scenarios and their interfaces, which includes a set of valid messages and states. The underlying \textit{access layer} defines the format of application data.

\paragraph{Multi-hop forwarding} To span multi-hop topologies, BT mesh nodes use a \textit{relay feature} which floods packets to omit routing complexity. This mechanism is known as \textit{managed flooding} since it includes methods for avoiding duplicates and loops. Its two main concepts are a hop limit counter per packet as well as a network message cache.

The \textit{network message cache} includes a list of messages recently seen by a node. If a message is already known, it is ignored and not forwarded.
This cache is mandatory for all nodes to prevent redundant forwarding and thereby to reduce network load on the mesh. The Bluetooth mesh standard does not require to cache the whole packet payload and leaves this open to the specific implementation. Hence, content caching is not originally supported. We will explore the differences to caching in ICN in the following sections.

\subsection{Data networking} Data exchange is organized using a publish-subscribe approach. Data is announced via a publish broadcast message to any of the addresses mentioned above. A publish can either be an unsolicited message initiated by any node in the network or it is sent as a reply to a previously received message. Replies are sent back to the BT source address of the requester. It should be noted that these unicast messages are mapped to broadcast advertisements on the lower layer.
Reception of messages is done by subscribing to addresses.
\textit{Models} specify the basic functionality of mesh nodes. Thereby, a node can act as client, server, or control node that implements both functionalities and adds the intermediate control logic.

Like the entire BT family, BT mesh is designed for local, isolated networks as there is no  way to transmit packets via transparent gateways to the Internet. Wide area communication can only be reached with the help of application layer gateways that translate the BT mesh communication Models into corresponding application protocols.

\subsection{Support for constrained environments}

Constrained nodes usually run on batteries, which quickly drain by the continuous channel scanning of Bluetooth mesh.
To foster sleep, Bluetooth mesh introduces so-called \emph{Friend nodes} that act as  proxies for the constrained devices.
Any constrained node is assigned to exactly one friend, but a friend may cover multiple constrained nodes.

After establishing this relation, the more powerful node maintains a message queue for each partner. Whenever the low-power node wakes up, it requests cached packets from its friend node.

\subsection{Security}
All communication is secured by authenticating and encrypting every packet using pre-shared, symmetric keys.
In a packet, all data above the network layer is encrypted using an \textit{application key}, which is shared among nodes belonging to a defined application domain (\ie all lights and switches on a building floor).
The complete packet, including the already encrypted application payload, is further encrypted using a \textit{network key}.
Network keys are shared among all nodes in a subnet, allowing them to authenticate and forward packets without disclosing any application data.

All keys are initially assigned during node provisioning and can be updated during runtime using specified update procedures defined by the BT mesh standard.

To prevent a simple passive eavesdropper from tracking nodes, BT mesh applies obfuscation techniques to key packet identifiers such as source addresses and sequence numbers.
In addition, packets are equipped with monotonic sequence numbers to protect against replay attacks, as well as an \textit{IV Index}.
This is a shared number between network members, which acts as entropy source for nonce generation and can be updated in case of state compromise.
Updating the index triggers a predefined procedure of distributed control.

\section{Bluetooth mesh vs ICN}\label{sec:icn-ble}
\begin{figure}
  \centering
  \input{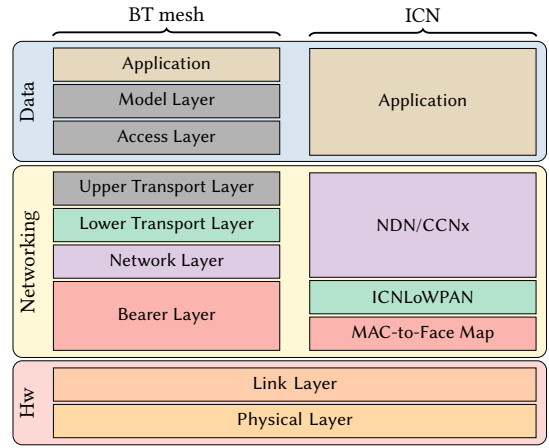}

\begin{tikzpicture}
  \pgfdeclarelayer{background}
  \pgfdeclarelayer{foreground}
  \pgfsetlayers{background,main,foreground}

  % BT mesh stack
  \node[boxshare, fill=Pastel1-E] (pl) {Physical Layer};
  \node[boxshare, fill=Pastel2-B, yshift=1pt] (ll) at (pl.north east) {Link Layer};
  %\node[box, fill=Pastel2-F, yshift=1pt, minimum height=0.50cm] (hci) at (ll.north west) {Host Controller Interface};
  \node[box2, fill=Pastel1-A, yshift=.25cm, minimum height=1cm+1pt] (br) at (ll.north west) {Bearer Layer};
  \node[box2, fill=Pastel1-D, yshift=1pt] (net) at (br.north west) {Network Layer};
  \node[box2, fill=Pastel2-A, yshift=1pt] (transl) at (net.north west) {Lower Transport Layer};
  \node[box2, fill=Set2-H, yshift=1pt] (transh) at (transl.north west) {Upper Transport Layer};
  \node[box2, fill=Set2-H, yshift=.25cm] (acc) at (transh.north west) {Access Layer};
  \node[box2, fill=Set2-H, yshift=1pt] (model) at (acc.north west) {Model Layer};
  \node[box2, fill=Pastel1-G, yshift=1pt, anchor=south west, ] (appble) at (model.north west) {Application};

  % ICN stack
  \node[box2, fill=Pastel1-A, yshift=.25cm, anchor=south east] (mac) at (ll.north east) {MAC-to-Face Map};
  \node[box2, fill=Pastel2-A, yshift=1pt, anchor=south east] (icnlo) at (mac.north east) {ICNLoWPAN};

  \node[box2, fill=Pastel1-D, yshift=1pt, anchor=south east, minimum height=1.5cm+2pt] (ndn) at (icnlo.north east) {NDN/CCNx};
  \node[box2, fill=Pastel1-G, yshift=.25cm, anchor=south west, minimum height=1.5cm+3pt] (appndn) at (ndn.north west) {Application};

  % arrow lower trans and icnlowpan
  % \draw[<->] (transl.east) -- node[] {} (icnlo.west);

  %\node[smallboxleft, fill=Pastel1-D, yshift=1pt] (gap) at (br.north west) {GAP};
  %\node[smallboxleft, fill=Pastel1-D, yshift=1pt] (att) at (gap.north west) {ATT};
  %\node[smallboxleft, fill=Pastel1-D, yshift=1pt] (gatt) at (att.north west) {GATT};
  %\node[smallboxleft, fill=Pastel1-H, yshift=1pt] (icnss) at (gatt.north west) {ICNSS};
  %\node[largeboxright, fill=Pastel2-A, yshift=1pt] (icnlowpan) at (br.north east) {ICN-LoWPAN};
  %\node[largeboxright, fill=Pastel1-B, yshift=1pt] (ndnccn) at (icnlowpan.north east) {NDN / CCNx};
  %\node[largeboxright, fill=Pastel1-G, yshift=1pt] (app) at (ndnccn.north east) {Application};

  % caption on top
  \draw[thick,decorate,decoration={brace,raise=5pt,amplitude=3pt}] (appble.north west) -- node[align=center, yshift=.5cm] {BT mesh} (appble.north east);
  \draw[thick,decorate,decoration={brace,raise=5pt,amplitude=3pt}] (appndn.north west) -- node[align=center, yshift=.5cm] {ICN} (appndn.north east);

  % background
  \begin{pgfonlayer}{background}
    \node[draw,fill=Pastel1-A!50,rounded corners,inner sep=0pt,fit={([xshift=-0.6cm,yshift=-0.1cm]pl.south west) ([xshift=0.15cm,yshift=0.1cm]ll.north east)},label={[rotate=90,anchor=north]left:{Hw}}] {};
    %\draw[fill=Pastel1-A!50,rounded corners] ([xshift=-.6cm,yshift=0.1cm]ll.north west) rectangle ([xshift=0.15cm,yshift=-0.1cm]pl.south east) node[pos=0.25,anchor=center, rotate=90] {Hardware};
    \node[draw,fill=Pastel2-F!50,rounded corners,inner sep=0pt,fit={([xshift=-0.6cm,yshift=-0.1cm]br.south west) ([xshift=0.15cm,yshift=0.1cm]ndn.north east)},label={[rotate=90,anchor=north]left:{Networking}}] {};
    %\draw[fill=Pastel2-F!50,rounded corners] ([xshift=-.6cm,yshift=0.1cm]transh.north west) rectangle ([xshift=0.15cm,yshift=-0.1cm]mac.south east) node[pos=0,anchor=north east, rotate=90] {Networking};
    %\draw[fill=Pastel1-B!50,rounded corners] ([xshift=-.6cm,yshift=0.1cm]appble.north west) rectangle ([xshift=0.15cm,yshift=-0.1cm]appndn.south east) node[pos=0,anchor=north east, rotate=90] {Data};
    \node[draw,fill=Pastel1-B!50,rounded corners,inner sep=0pt,fit={([xshift=-0.6cm,yshift=-0.1cm]acc.south west) ([xshift=0.15cm,yshift=0.1cm]appndn.north east)},label={[rotate=90,anchor=north]left:{Data}}] {};
  \end{pgfonlayer}
\end{tikzpicture}
  \caption{Bluetooth mesh stack (left) vs our ICN stack (right).}%
  \label{fig:blemesh_ndn}
\end{figure}

\subsection{Stack overview}
BT mesh specifies a full vertical stack, where the network layer is only one part.
In contrast, ICN defines a network architecture relying on supporting protocols on the layers above and below for successful deployment.
Figure~\ref{fig:blemesh_ndn} compares overall views of the BT mesh design with an ICN stack as we propose for constrained IoT networks.

Both stacks can utilize the same link technology, but BT mesh only defines bearers that are tied to BLE radios where ICN is not restricted to those.
For multi-hop communication BT mesh is further restricted to the broadcast-based advertising bearer.
The ICN stack may use a face to link mapping selecting the outgoing unicast channel (\ie, Layer-2 unicast destination address) according to the network-layer routing or forwarding decisions.

On top, the ICNLoWPAN~\cite{gksw-innlp-19} convergence layer implements segmentation and compression to adapt and reduce data to be sent over low-power links.
In contrast, BT mesh does not compress any headers and implements segmentation  on top of the network layer in the transport layers, since its link properties are predefined and no flexible adaption is needed.
Unlike NDN, networking in BT mesh defines a stacked transport layer that covers segmentation,
reassembly, but also control traffic.

The ties to BLE and the vertical integration make BT mesh a locally confined, monolithic technology, whereas ICN is open w.r.t. various link layers, applications, and wide-area deployment.

\subsection{Networking}

\paragraph{Routing and forwarding}
BT mesh only supports a single network interface.
Furthermore, it does not support multiple faces over this interface as the used link layer does not support destination addresses.
Multi-hop and multi-path topologies are built without routing by simply forwarding all packets over the broadcast channel of this interface.
Although flooding follows a managed approach, it leads to substantial resource consumption for all nodes in reach.
In contrast, NDN  guides Interest packets according to a Forwarding Information Base (FIB), which can be populated by a variety of routing protocols.
In addition,  multiple requests (solicited publishes in BT) to the same resource can be aggregated in the Pending Interest Table (PIT) and drastically reduce the number of packet traversals in a multi-hop environment \cite{rr-dicpc-12,gklp-ncmcm-18}.
On the other hand, BT mesh natively supports group requests, as all responses are returned to the initiator.
Data aggregation in ICN is not yet visible as a standard service.
A proposal for multi-source data retrieval in ICN has been published in \cite{acm-msdri-14}.

\paragraph{Reliable transport}
BT mesh supports transmission of unacknowledged and acknowledged messages, where the latter requires each
subscriber to send a response. The message type in use is defined by the application \textit{model} and can be adjusted
during runtime via control models. Publish nodes are allowed to retransmit messages until a new message of the same model
is published, which stops ongoing retransmissions. In contrast, NDN implements a best-effort hop traversal by retransmitting pending Interests on time triggers until data returns, a NACK is received, or PIT entries time out.

The BT mesh standard advises against the use of acknowledged messages in multicast scenarios, as the number
of expected responses is unknown. This burdens application logic with implementing acknowledgements and
timeout.
The flooding nature of BT mesh plus its retransmission capabilities can lead to high resource demands.
The effects and a collection of counter-measures were presented in \cite{dg-blemn-17,gmis-pable-16,klj-bwmnp-15}.

\paragraph{Medium Access Control (MAC) layer adaptation}
Although BLE defines reliable point-to-point links that
involve time division and frequency hopping, BT mesh simply
broadcasts on the advertising channels, which discards the original reliability features of BLE.
NDN, on the other hand, does not natively define a mapping of faces to the MAC layer. In typical
deployments however, faces are often mapped directly to the broadcast address similarly,
or to the unicast MAC address. While broadcasting enables path diversity with no
additional routing overhead, it increases resource consumption, whereas a mapping of faces to
unicast traffic saves resources but comes to the cost of neighbor management. These effects have been
 analyzed in \cite{kgshw-nnmam-17}.

\paragraph{Application layer adaptation}
In comparison to BT mesh that specifies a
data \textit{model} for different device types, such as light sensors and dimmers, the main
requirement to forwarding data to the application layer in NDN is a matching PIT entry.
Respectively, there is no strict requirement on the format of names. This
shifts complexity to the namespace but on the other hand, flexibly widens the field of possible
applications. Introducing a new traffic type in BT mesh requires
implementation of a new model. By restricting applications to models, BT mesh reduces its areas of use.

Regarding its {\bf overall networking capabilities}, BT mesh starts from utmost flexibility on the lower layers by broadcasting. Up the stack it narrows down by restricting networking functions and---even further---the application interfaces. NDN follows an opposite approach. It reduces lower layers to pipes behind faces that were initially understood to interface point-to-point links, but remains open to various routing and forwarding strategies and leaves applications with a simple, yet fully flexible name-based network access.

\subsection{Names and publish-subscribe}

BT mesh uses (group) addresses to label data streams.
NDN acts upon hierarchical tokenized names assigned to data entities.
The semantic representation in NDN gives several degrees of freedom in terms of network configuration and application design whereas the addressing scheme in BT mesh is mostly static and normally configured by manual provisioning of a node before deployment.
In both cases, however, data is addressed and not an endpoint, which makes BT mesh content-centric like NDN.

Publish-subscribe services require a rendezvous function to bring together publishers with subscribers, which are decoupled otherwise.
The common way to implement rendezvous is a well-known broker that mediates between peers.
Some ICN flavors such as PSIRP/Pursuit and related \cite{lvt-psipp-10, cpw-cpsni-11} natively follow such a publish-subscribe paradigm \cite{xvsft-sinr-14} by foreseeing a broker function as part of the network service architecture.
NDN and CCNx are based on a request-response mechanism without a push option for data. In contrast, publish is commonly performed as a push operation and subscription as a state that awaits it, which led to workarounds and discussions in previous work  \cite{cajfr-cecop-11,acim-ndnia-14,gksw-hrrpi-18}.
Nonetheless, support publish-subscribe operations can be realized without protocol modifications using NDN primitives \cite{gksw-hrrpi-18,szabw-pcbms-18,lsawz-binds-18,n-llbsn-19}.
Here, rendezvous is provided by names or prefixes, which are matched by some content repository or cache that is well known to the routing system.

In BT mesh all content is distributed via publish-subscribe.
The approach in BT mesh does not employ a broker, but advertises content items to group addresses. Group advertisements are filtered by the upper layers so that only subscribers of a specific group receive them.
Conversely,  publishing in BT mesh can be a solicited operation, which then resembles
the request-response scheme of ICN. If a publish hits a subscriber, it can trigger a
publish being sent back to the initiator. This does not prevent broadcasting of publish requests on the link layer, though.

By relying on broadcasts, BT mesh remains independent of any  infrastructure entity such as a service broker. This feature, however, is achieved by significantly straining the network. The latter imposes severe scaling limits, as we will demonstrate in our experiments.

\subsection{Low power assistance and caching}

The low-power friendship relation is the feature of BT to integrate very constrained, battery driven nodes.
As a default, this role is not defined in ICN, but there exists work on investigating this topic \cite{xzlz-ardaw-18}.
Caching principles are present in both stacks.

In BT mesh, friend nodes allocate caches for each constrained node that has established a relation before going to sleep.
When a constrained node awakes, it notifies its friend which then sends all cached packets.
On relay nodes, different caches are responsible for identifying previously seen packets.
They act as a forwarding filters for managed flooding and do not serve data---in contrast to the ICN approach of distributed in-network caching.

It is noteworthy that BT mesh nodes can always only have one cache available at its friend. This cache drops entries after their first retrieval, which counters a re-use of cached data. In contrast, NDN can cache any data packet at any node for access throughout the
network and the cache replacement strategy is open to implementers and deployment.

%%% Local Variables: mode: latex TeX-master: "main" End:

\section{Evaluations}\label{sec:eval}

\subsection{Experiment Setup}

\paragraph{Hardware Platform}
We deploy our experiments on common low-end IoT hardware based on ARM Cortex-M providing 64~kB RAM and 512~kB flash.
For the BT mesh measurements, we use nRF52dk development kits from Nordic Semiconductor featuring BLE-compliant Cortex-M4F SoC (nRF52832) running at 64~MHz.
Our NDN measurements run on iotlab-m3 nodes featuring a Cortex-M3 SoC (STM32F1) at 72~MHz.
They carry an 802.15.4 radio (Atmel AT86\-RF231) that provides basic MAC features such as address filtering in hardware.
All nodes are part of the Saclay site of the FIT IoT-LAB testbed.

We use different hardware platforms to approach real deployment as close as possible.
The differences between both CPUs (instruction set, core clock frequency) are neglectable and do not have a noticeable impact on our measurements.
The radios on the NDN and BT mesh platforms, however, differ in some aspects.
The Nordic radio in BT mesh mode uses the default BLE bit rate of 1~Mbit/s and does not support carrier sense multiple access (CSMA) and automatic repeat request (ARQ).
On the other hand, the Atmel radio runs at 250~kbit/s and features CSMA and ARQ.

To check for consistency, we performed all measurements related to NDN on both platforms.
The traffic load results are equal.
The content arrival time measurements show similar success rates but vary in the time to completion, which roughly corresponds to the differences in radio bit rates.

We argue that the CSMA and ARQ support of the iotlab-m3 platform comes much closer to a real-world production deployment than the proprietary mode of the Nordic radio.
For this reason, all NDN measurements presented in the following have been done on the iotlab-m3 boards.

\paragraph{Software Setup}
Our experiments are based on the RIOT~\cite{bghkl-rosos-18} operating system version
\texttt{2019.04}. To analyse BT mesh, we use the NimBLE stack.\footnote{\url{https://github.com/apache/mynewt-nimble}}
Our NDN measurements are based on CCN-lite~\cite{ccn-lite}.
Both network stacks are available as third-party packages in RIOT.
In the Saclay testbed all nodes are in wireless range of each other.
To employ multi-hop topologies, we apply MAC address filtering on the nodes.
Although this enforces a virtual topology on the network layer, it does not prevent two nodes, that topologically do not see each other, to cause interference on the physical layer.

\paragraph{Configuration}
We align parameters between experiments where possible to follow a consistent setup.
On the link layer, we used the default parameters given by NimBLE and adapted CCN-lite to apply comparable behavior.
BT mesh always transmits every packet 5~times per channel (5 advertising events) with an interval of 20~ms.
For IEEE~802.15.4 we enable link layer acknowledgments with a maximum of 4~retransmissions.
Additionally, we run CCN-lite with 4 network layer Interest retransmissions, a retry interval of 1~second, and a timeout value of 10~seconds.
The content store (CS) is limited to 30 entries.

\paragraph{Scenarios}
We deploy \one 10~nodes in a full mesh to analyze single-hop scenarios, and \two the same nodes along a line to
evaluate multi-hop use cases. The ICN stack configures FIBs with faces that map to unicast MAC addresses of their neighbors.
The BT mesh stack uses a group address on the network layer but will broadcast on the link layer.
For each topology, we configure two different traffic patterns:

\begin{description}
    \item[Many to one] One node acts as data sink (consumer) and all other nodes are data sources (producers). In ICN, the consumer requests data from all producers iteratively, which reflects common ICN scenarios. In BT mesh, every producer node publishes data to a group address to which the sink is subscribed.
    In both scenarios, each producer publishes 100~content items with a publishing rate of one item every 5~s $\pm$ 2.5~s of jitter.
    This publishing rate is limited by the publisher node's system resources.

    \item[One to many] One node acts as producer and all other nodes are data consumers. In ICN, all consumers request data 'simultaneously'. In BT mesh, the producer publishes data to a group address. In both cases,
        100~content items are published by a producer with a publishing rate of one item every 1s $\pm$ 0.5~s of jitter.
\end{description}

\iffalse
\begin{table}[]
\caption{BT mesh vs ICN success rates.}%
\label{tab:success_rates}
\begin{tabular}{r  cc cc}
\toprule
        & \multicolumn{2}{c}{One-to-many} & \multicolumn{2}{c}{Many-to-one} \\
        \cmidrule(r){2-3} \cmidrule(l){4-5}
        &single-hop&multi-hop&single-hop  &multi-hop\\ \midrule
BT mesh &100~\%    & 100~\%  &88,4~\%      &89~\%\\
ICN     &100~\%    &100~\%   &100~\%      &100~\%
\\
\bottomrule
\end{tabular}
\end{table}
\fi

\subsection{Results}
\begin{figure}
    \begin{subfigure}[l]{.23\textwidth}
        \includegraphics[width=1\textwidth]{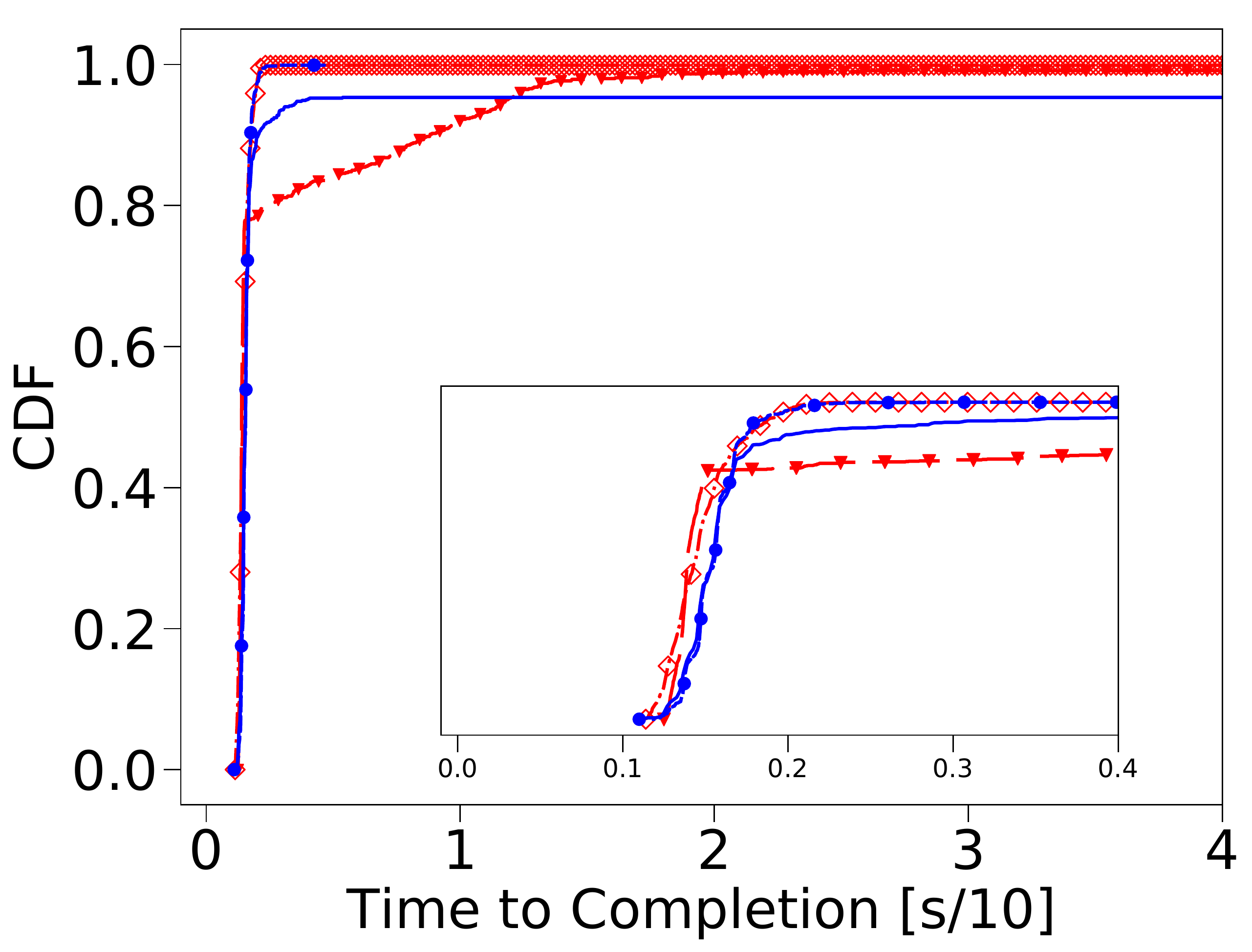}
        \subcaption{Single-hop}
        \label{fig:ttc_plots_a}
    \end{subfigure}
    \begin{subfigure}[r]{0.23\textwidth}
        \includegraphics[width=1\textwidth]{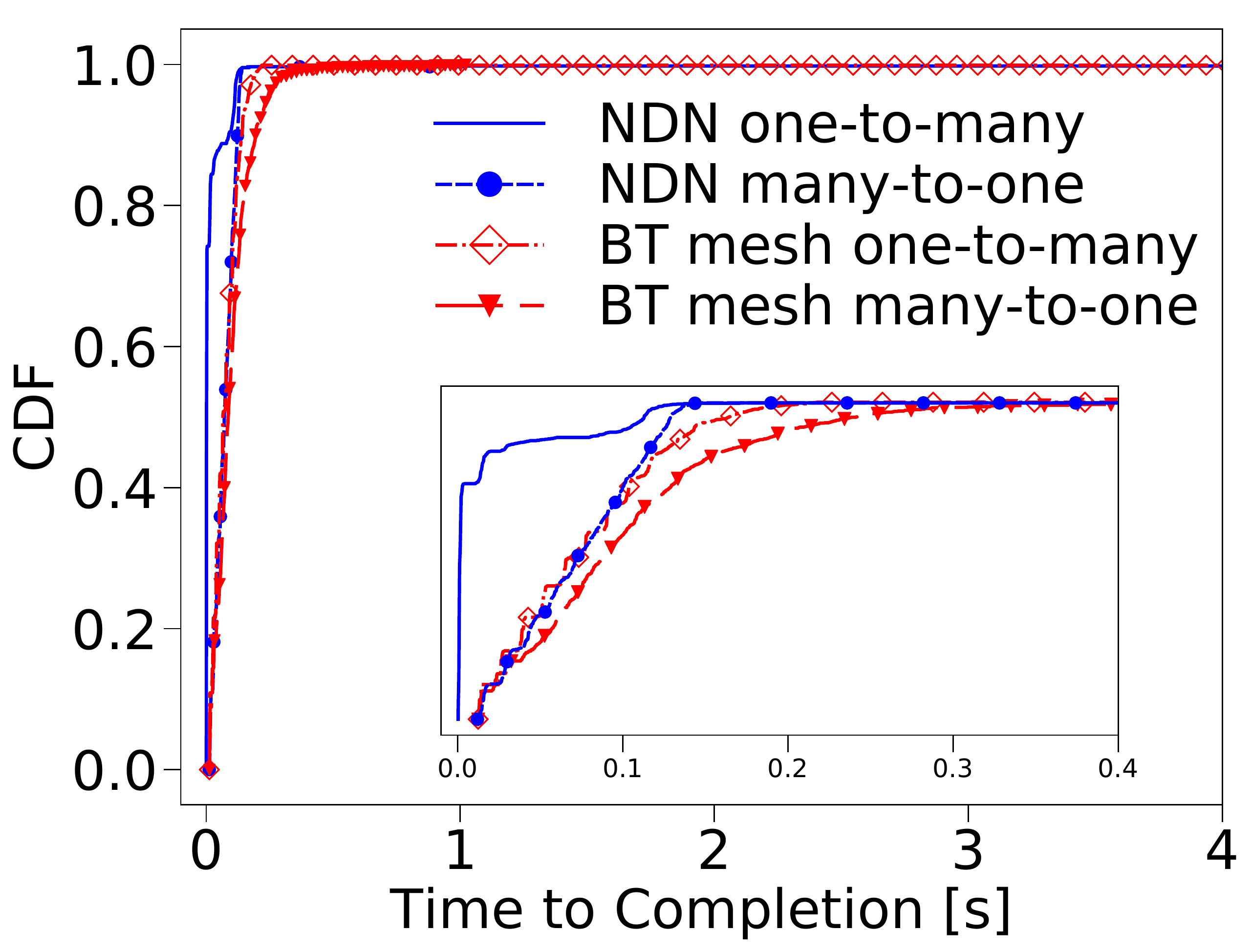}
        \subcaption{Multi-hop}
        \label{fig:ttc_plots_b}
    \end{subfigure}
    \caption{Time to content arrival.}
    \label{fig:ttc_plots}
\end{figure}

The average content delivery rate for each measurement setup is 100\%.
Network performance, however, is more subtle.

\paragraph{Content Arrival Time (Single-hop)}
We first measure the time to content arrival for each data packet sent from a publisher to a consumer.
Figure~\ref{fig:ttc_plots} shows the cumulative frequency of all events for all configurations.

In the single-hop scenario (cf., Figure~\ref{fig:ttc_plots_a}), NDN and BT mesh exhibit almost the same behavior for the many-to-one and one-to-many scenario respectively.
Nearly 80\% of the contents arrive during the first 15~ms.
Even though NDN sends two packets per content (Interest and data) differences compared to BT mesh are negligible.
The data delivery time is only limited by the local link layer transmission delay because no competing packets occur.

In the one-to-many case of NDN, the results change slightly.
Most noticeable is that the content arrival time exceeds 20~ms for 5\% of the published content.
In fact, very few content items need up to 1~s (not shown).
In this scenario multiple consumers request content independently and in parallel, which results in some packet loss of Interest messages that are (successfully) retransmitted after 1~s.

In the many-to-one scenario of BT mesh, $\approx$80\% of the content items arrive also within the first 15~ms.
The remaining items need up to 120~ms to be successfully delivered.
The reason for this relies in the broadcast nature of BT mesh and the simple reliability scheme which requires uninformed retransmissions.
In our case every node sends each content item five times with a delay of 20~ms, explaining the linear slope converging around 120ms.
Having retransmissions without packet loss does not pay off but leads to increased collisions, even in small networks.

\begin{figure*}
    \begin{subfigure}[l]{.24\textwidth}
        \includegraphics[width=1\textwidth]{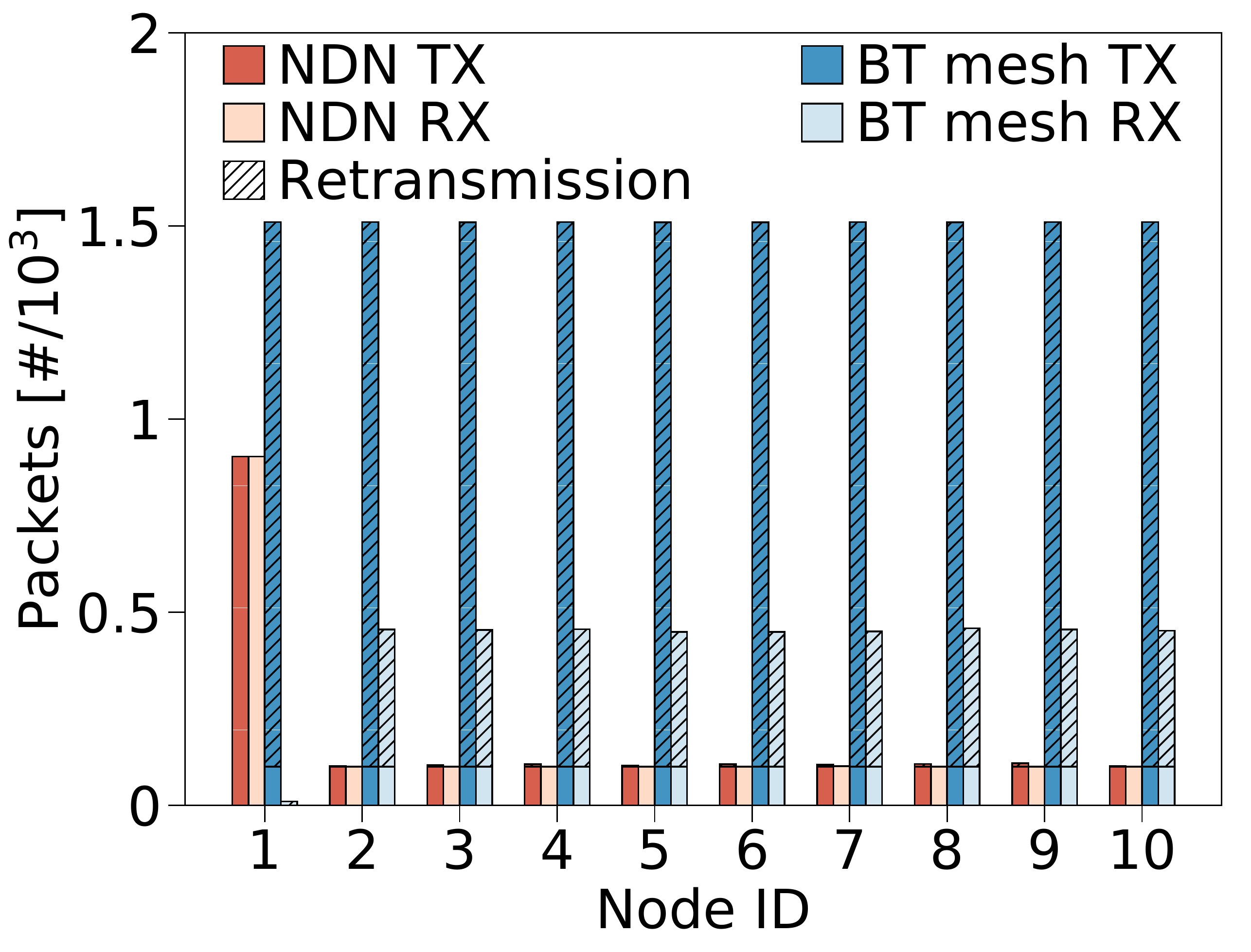}
        \subcaption{Single-hop, one to many.}
        \label{fig:txrx_cnt_a}
    \end{subfigure}
    \begin{subfigure}[r]{0.24\textwidth}
        \includegraphics[width=1\textwidth]{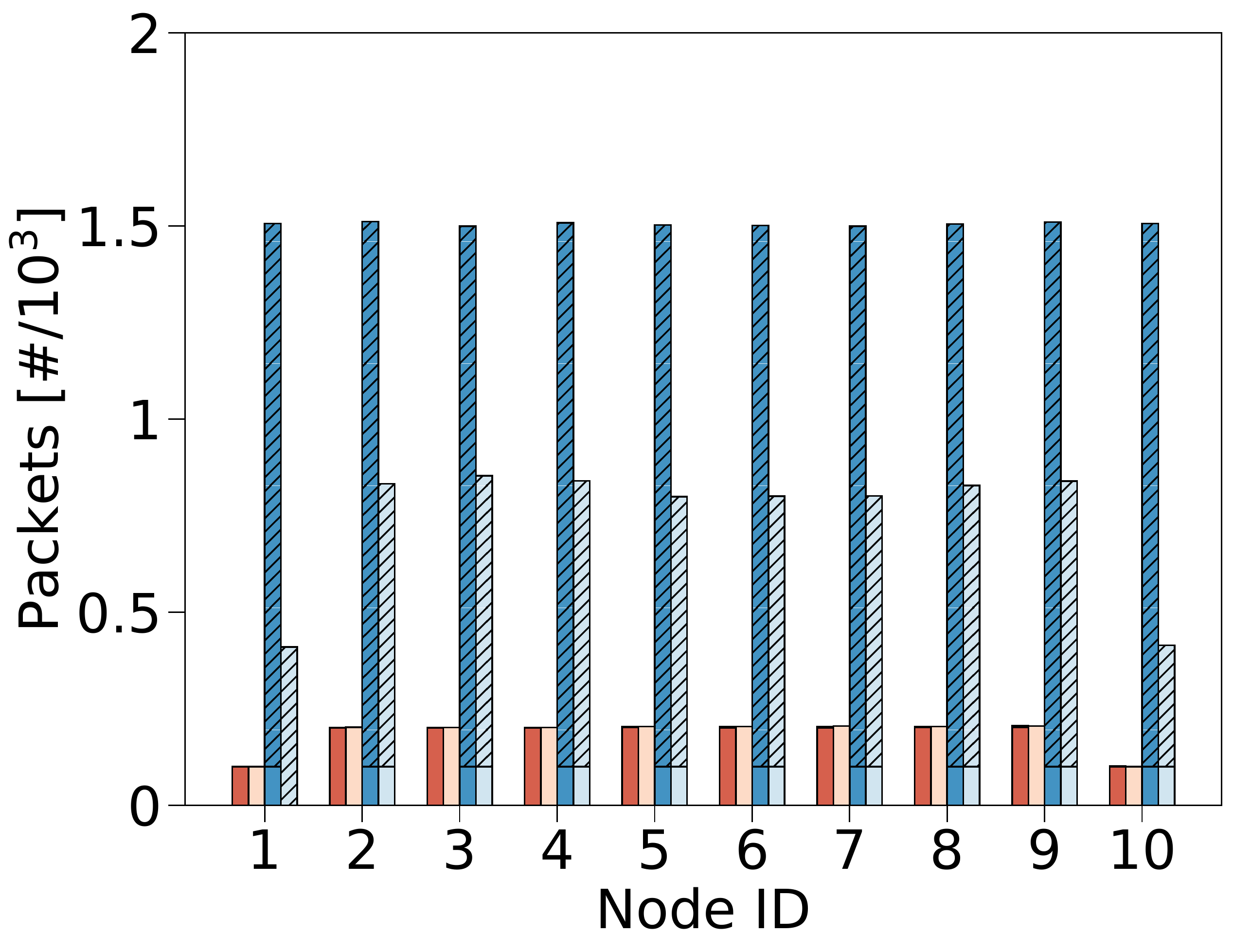}
        \subcaption{Multi-hop, one to many.}
        \label{fig:txrx_cnt_b}
    \end{subfigure}
    \begin{subfigure}[r]{0.24\textwidth}
        \includegraphics[width=1\textwidth]{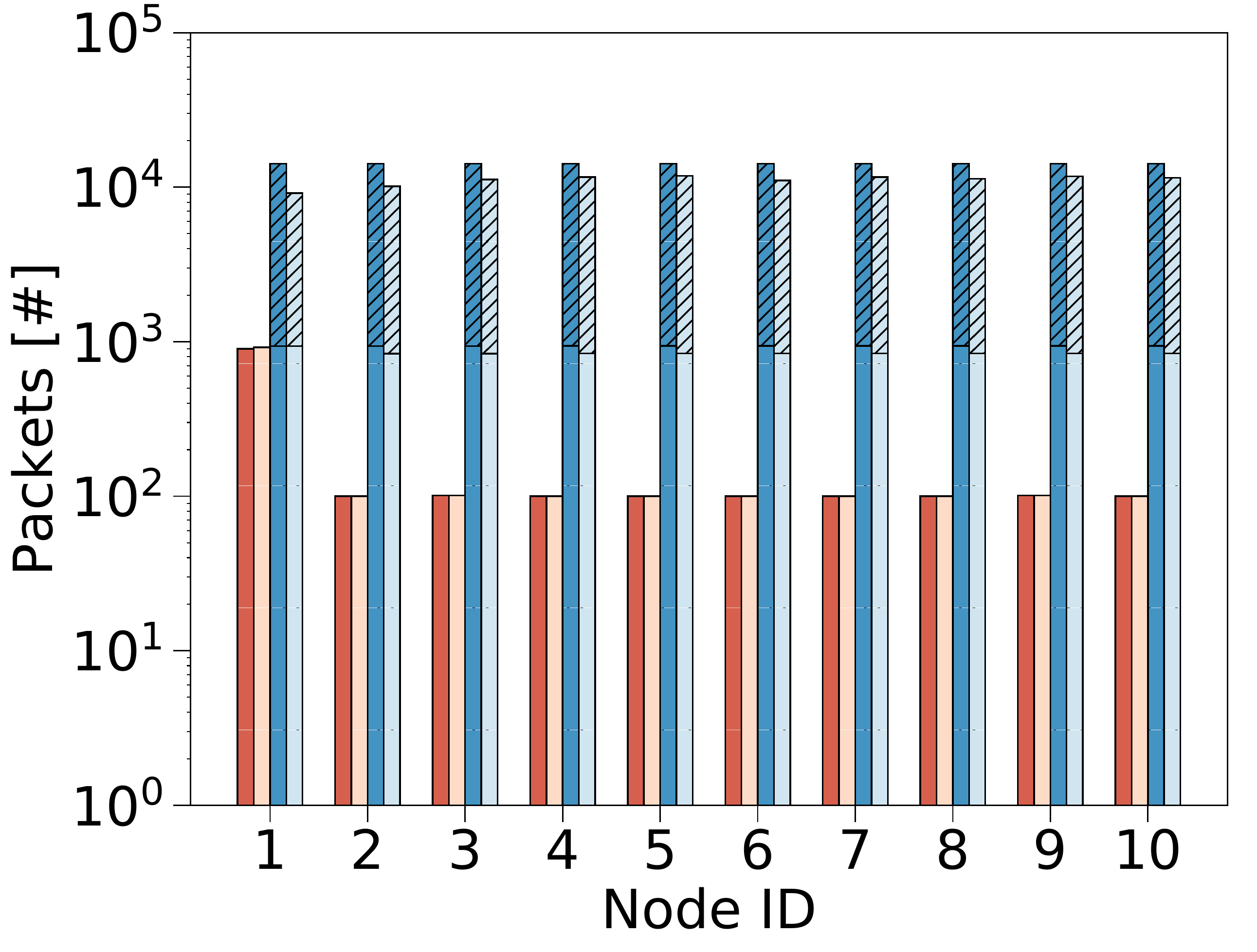}
        \subcaption{Single-hop, many to one.}
        \label{fig:txrx_cnt_c}
    \end{subfigure}
    \begin{subfigure}[r]{0.24\textwidth}
        \includegraphics[width=1\textwidth]{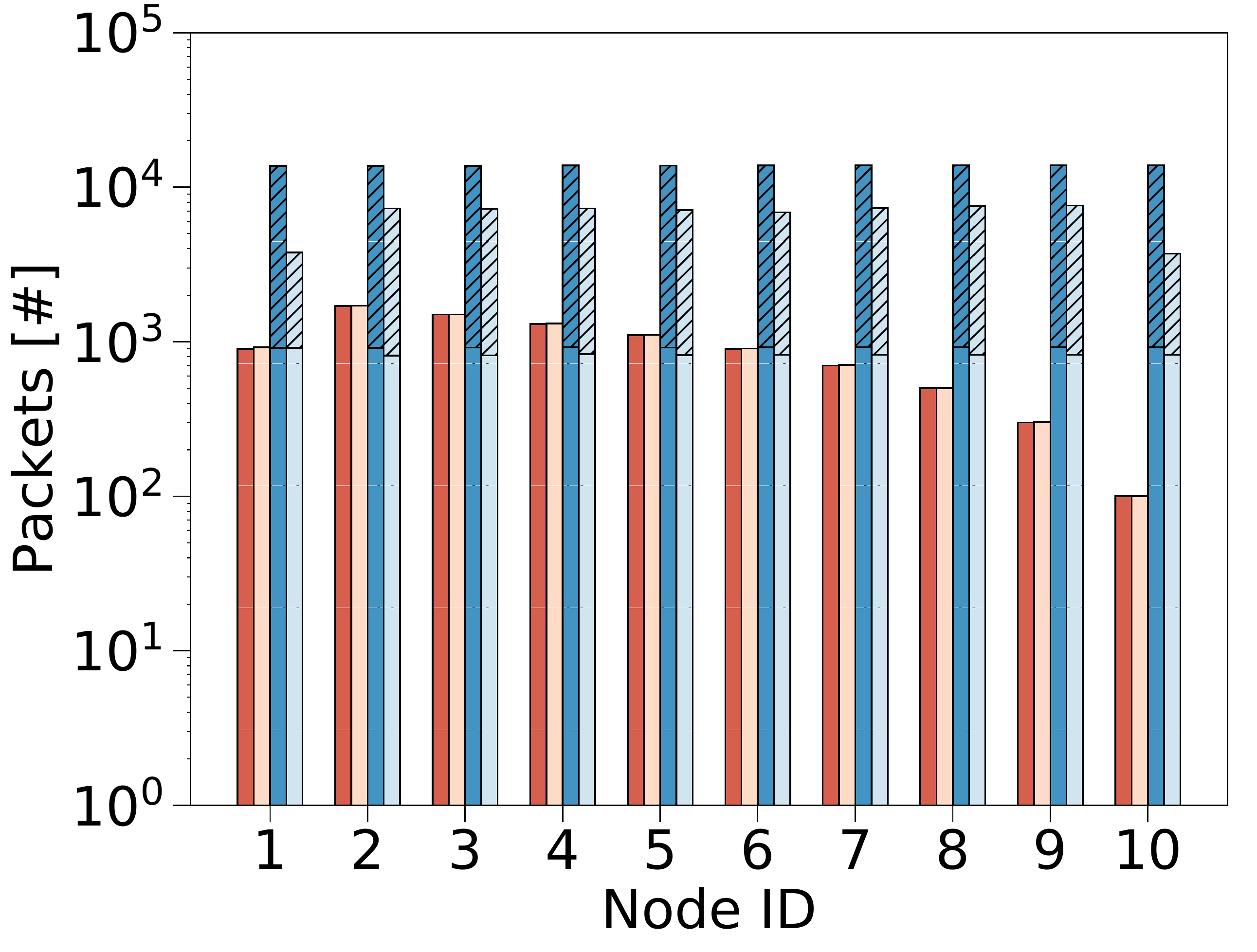}
        \subcaption{Multi-hop, many to one.}
        \label{fig:txrx_cnt_d}
    \end{subfigure}
    \caption{Traffic load for BT mesh and NDN.}
    \label{fig:txrx_cnt}
\end{figure*}

\paragraph{Content Arrival Time (Multi-hop)}
Multi-hop communication delays content delivery by one order of magnitude, see Figure~\ref{fig:ttc_plots_b}.
The only remarkable exception of a significant performance decrease is NDN in the one-to-many scenario.
Here, 75\% of the content is delivered almost immediately ($\le$~1~ms) because of in-network caching and properties of the topology.
A line topology enforces that a content item quickly passes all (consumer) nodes, which will cache the content.
When consumers that did not request content before, send their Interest, they benefit from their local cache.
Then, the data delivery time is mainly limited by local system resources (RAM access \etc).

In the many-to-one and one-to-many use case of NDN and BT mesh respectively, data delivery does not significantly diverge, for similar reasons as in the single-hop scenario.
It is worth noting that caching in NDN becomes only effective in case of retransmissions, because a single consumer requests content in this scenario.
Surprisingly, in the many-to-one use case, BT mesh performs similar to the previous configuration, with less divergence compared to single-hop.
We attribute these results to data submission on different channels.
Providing non-overlapping channels between two peers reduces collisions in particular in multi-hop scenarios.

\paragraph{Traffic Load}
We now analyze the traffic load per node, differentiating the original network traffic and retransmissions in place (cf., Figure~\ref{fig:txrx_cnt}).
NDN clearly outperforms BT~mesh.
In contrast to BT mesh, NDN reduces the overall traffic load due to improved error handling, \ie packets on the network layer are retransmitted only when needed, and caching.
Caching effects are nicely visible in the line topology (cf., Figure~\ref{fig:txrx_cnt_b}), and comply with our discussions about the content delivery time.

BT mesh experiences many more packets per node.
Any new content item is replicated a priori, leading to at least 1500 sent packets in our topology.
Furthermore, content is relayed by each node that receives it, leading to additional 1500 sent packets by each relay, and increased receive overhead.
Note that each BT mesh node is only listening to 1 of the 3 advertising channels at any time.

In the single producer scenarios (Fig \ref{fig:txrx_cnt_a} and \ref{fig:txrx_cnt_b}), the amount of transmitted packets is close to equal for all BT mesh nodes, as all packets are evenly relayed throughout the network, independent of its topology.
For the single producer, multi-hop setup, NDN demonstrates the advantage of the in-network caching by a reduced network load at the producer node.

Looking at the multi producer scenarios (Fig. \ref{fig:txrx_cnt_c} and \ref{fig:txrx_cnt_d}), the effect of BT mesh's static link layer retransmissions is even more distinctive.
The number of transmitted and received network packets is again roughly equal for the single- and multi-hop setups, but their absolute numbers are multiplied due to the increase in produced content items.
In contrast, NDN shows comparable behavior to the single producer scenarios, as here only the source and destination of interest and content packets is turned around.

In general, the static link layer retransmission of BT mesh amplifies network load in all of the measured scenarios.

%%% Local Variables:
%%% mode: latex
%%% TeX-master: "main"
%%% End:

\section{BT-ICN: ICN perspective for BLE}\label{sec:perspect}

In the following, we discuss two perspectives for joining the best properties out of both worlds: (i) building BT mesh-like properties by using BLE as a reliable transport for ICN, and (ii) implement BT mesh's friend relationship applying ICN principles.

\paragraph{ICN-over-BLE}
BLE's connection-oriented mode provides a resilient link based on time sliced channel hopping (TSCH).
Running ICN over such BLE links, while tightly coupling ICN routing to BLE connection management, promises to be efficient and reliable in terms of network resource usage.
Such a setup can potentially deliver similar service properties as BT mesh, while taking full advantage of both, caching and routing capabilities of ICN, as well as the reliable BLE link layer.

\paragraph{Friend role using ICN principles}
Following, we propose an implementation of the low-power friend feature in ICN to demonstrate the similarity of the friend feature and the principles of NDN.

A friend node that requests content from its sleepy constrained node may establish PIT state on both participants with a \textit{long-lived Interest} at a point in time, with a PIT timeout longer than the node's sleep cycle.
The node may reply to the Interest whenever it wakes up.
Alternatively, a method like HoPP~\cite{gksw-hrrpi-18} may be used to publish a content to the friend node in a timely manner.

If a constrained node wants to request content, it may simply state an Interest to its friend.
If the content is not replied shortly, the low-power node may decide to repeat the Interest after some time, with a high probability that the request is now satisfied from the friend's content cache.

In BT mesh, dedicated caches are maintained for each low-power node.
This is simplified by our approach, as the friend nodes need simply to deploy vanilla ICN caching.
\section{Related Work}\label{sec:related-work}

The use of BLE and BT mesh in low-power environments has been analyzed from
various perspectives:
\one IETF standard solutions for IPv6 over connection oriented BLE networks have been
released in \cite{RFC-7668}, extended, and analyzed in \cite{llkb-sarb-16,sbzr-befpi-17}.
There are ongoing efforts to standardize mesh topologies for IPv6 traffic over BLE \cite{draft-ietf-6lo-blemesh-05}.
\two Initial performance analysis of plain BT mesh networks were presented in \cite{brsh-bmsoe-18,dsg-bmecm-19}
and a collection of academic solutions deal with improvements on the routing mechanism, which
are summarized in \cite{dg-blemn-17}.
\three Few work deals with Bluetooth in the context of ICNs. The
authors of \cite{s-nfbea-13} introduce a Bluetooth convergence layer for NetInf \cite{a-snad-10}
and provide an implementation for Android devices. \cite{ms-moses-14} bases on the this implementation and focuses
an opportunistic networking platform for content dissemination. In \cite{bils-smmbl-15}, the NDN
approach is applied for BT mesh networks to aggregate distributed databases.

Experimental evaluations of ICN in the constrained IoT employing the large FIT IoT testbed have started in 2014 \cite{bmhsw-icnie-14} and since then established a tradition of reality-checks in this rather unpredictable domain.

%%% Local Variables:
%%% mode: latex
%%% TeX-master: "main"
%%% End:

\section{Conclusions and Outlook}\label{sec:c+o}

Starting from the common claim that Bluetooth mesh implements ICN principles, we conducted a detailed comparative analysis.
Our theoretical and experimental study, however, revealed that there is no evidence for this statement.
BT mesh performs flooding without content caching, leading to degraded network performance, while ICN principles still need to be added in future work.

The holistic nature of BT mesh, providing a full vertical stack and covering all aspects of a nodes lifetime, makes it a production-ready solution.
While there exists a large number of candidate solutions that can be combined with ICN-based networking to achieve a similar set of functionality, some parts are not covered by proven go-to solutions (\ie provisioning and managing of security artifacts, managing of large-scale namespaces).

From a network perspective, BT mesh proved suitable for strictly local appliances (\ie smart lighting), provided the installation size does not grow too large and not too many neighbors use this technology simultaneously.
This is a significant limitation at the core of this technology.
ICN on the other hand is designed specifically to cope with large-scale deployments and we showed that NDN can utilize network resources more efficiently than BT mesh.

%%% Local Variables:
%%% mode: latex
%%% TeX-master: "main"
%%% End:

\subsection*{A Note on Reproducibility}
We fully support reproducible research~\cite{swgsc-terrc-17,bbfkp-dbgre-19} and perform all our experiments using open source software and an open access testbed.
Code and documentation is available on Github at\\ \url{https://github.com/5G-I3/ACM-ICN-2019-BLE}.

%%% Local Variables:
%%% mode: latex
%%% TeX-master: "main"
%%% End:

\subsection*{Acknowledgments}
We would like to thank the anonymous reviewers and our shepherd Alex Afanasyev for their valuable feedback.
This work was supported in parts by the German Federal Ministry of Education and Research (BMBF) within the 5G project \emph{I3} and the VIP+ project \emph{RAPstore}.

\balance
\bibliographystyle{ACM-Reference-Format}
\bibliography{own,rfcs,ids,ngi,iot}

\end{document}